\documentclass[aps,prl,twocolumn,showpacs,amsmath,amssymb]{revtex4}
\usepackage{graphicx}
\usepackage{dcolumn}
\usepackage{bm}
\begin{document}

\title{Coherent emission of atomic soliton pairs by Feshbach resonance tuning}
\author{Humberto Michinel$^1$, \'Angel Paredes$^1$, 
Mar\'{\i}a M. Valado$^{2,3}$ and David Feijoo$^1$}
\affiliation{$^1$\'Area de \'Optica, Facultade de Ciencias de Ourense,\\ 
Universidade de Vigo, As Lagoas s/n, Ourense, ES-32004 Spain.\\
$^2$Dipartamento di Fisica ``E. Fermi", Universit\`a di Pisa, Largo Pontecorvo 3, 56127 Pisa, Italy.\\
$^3$INO-CNR, Largo Pontecorvo 3, 56127 Pisa, Italy.
}

\begin{abstract}
We present two simple designs of matter-wave beam splitters in a trapped Bose-Einstein 
Condensate (BEC). In our scheme, identical pairs
of atomic solitons are produced by an adequate control --- in time and/or space ---
of the scattering length.
Our analysis is performed by numerical integration of the Gross-Pitaevskii equation and
supported by several analytic estimates. Our results show
that these devices can be implemented in the frame of current BEC experiments. The system
has potential applications for the construction of a soliton interferometer.

\end{abstract}


\maketitle


\section{I. Introduction}
One of the most promising research tracks in the field of Bose-Einstein Condensation (BEC) in gases \cite{Anderson95} 
is the design of interferometric devices using coherent matter waves \cite{Mewes97, RMP}.
The potential use of these novel atomic interferometers in new types of sensors \cite{sensors}, precision measurements \cite{precision} 
and in the study of gravitational effects \cite{gravitation}, among other areas, has become a very active topic and 
different schemes have been proposed, most of them based on the control of the atomic cloud by time-dependent
traps \cite{interferometer}.

Though the general principles of interferometry apply both to coherent light and matter-waves on equal
 footing, there are important distinctions
between optical and  atomic devices. Besides the different intrinsic nature of waves and the physical scales involved in both cases, one of the 
most remarkable gaps is that the design of accurate and controllable atomic beam splitters is much more difficult as compared to light systems --- it must be said, though, that it is possible to implement atomic
quantum interferometry without such a beam splitter \cite{hyllus}.
Moreover, once a pair of BEC pulses is produced, the atomic clouds will show a significant spreading due to internal repulsive interactions.  
This dilution of the waves may diminish dramatically the signal-to noise ratio of the device, precluding the detection of the effects under study,
specially if their traces are minimal as is the case of many gravitational and quantum interactions \cite{gross}. 

The use of atomic soliton sources \cite{solitons1,solitons2},
 which produce self-trapped matter wave packets that propagate undistorted, 
can be an interesting strategy to avoid the spreading and 
provide precision interferometric measurements. 
This possibility, already pointed out in \cite{solitons1}, has been recently pursued in
\cite{parker,castin,streltsov,billam,stoof,martin,helm}. 
It has been discussed how to engineer a coherent soliton beam splitter by  
manipulations of the external potential \cite{castin,streltsov,stoof,martin,helm}
or by making use of a Rabi coupling between two
atomic states of the particles in the sample \cite{billam}.

The goal of this work is to introduce new protocols for the control of atomic clouds
that can coherently produce soliton
pairs. The evolution is controlled by an appropriate tuning of the strength and sign of the 
inter-atomic interactions.
Our construction is similar to the simple pulsed atomic soliton laser
 first proposed by Carr and Brand in Ref. \cite{Carr04}, where simulations showed that a 
 train of robust matter pulses can be generated by the mechanism of 
modulational instability (MI). However, the device of \cite{Carr04} has  limited utility in the frame of interferometry due to the following drawbacks:  (i) the number 
of wave-packets generated cannot be controlled and they have different shapes (ii) the trap must be destroyed after outcoupling (iii) several 
pulses are always produced whereas in many applications single pair sources are of interest and (iv) the pulses will travel at different speeds 
once the trap is removed. 

In the following, we will show a simple 
mechanism that allows to overcome the previous limitations yielding atomic coherent
sources suitable for novel types of matter-wave interferometers. This is  interesting since the techniques for generating 
and controlling BEC with growing number of particles and their physical properties are nowadays well established and the current 
experimental challenges in the field face the design of practical devices \cite{chip}. 

Thus, in the present paper, we will demonstrate that pairs of counter-propagating 
atomic solitons can be emitted from a trapped BEC reservoir by an adequate tuning of the scattering length $a$. 
The idea is 
quite simple: if the atomic cloud is placed in the center of a shallow trap and $a$ is tuned to a large positive value, the atomic cloud 
will spread due to strong internal repulsive interactions. When the width of the cloud is large enough, the inter-atomic forces are switched 
from repulsive to attractive by means of Feshbach resonance tuning \cite{FB1, FB2}. Then, modulational instability yields soliton formation \cite{carrbrand2}. 
Once the symmetric pair of solitons is produced, the matter wave pulses leave the trap with opposite velocities. The key point is that {\em an 
adequate tuning of the scattering length allows full control over the splitting process}.
We will show that this idea can be accomplished by modulating $a$ either in time or in space.

 An atomic Michelson interferometer configuration can be easily 
implemented by the addition of a wide parabolic potential that forces the solitons to return and interfere. As we will show in the rest of this work, 
this simple device can be straightforwardly built in the frame of current experiments with ultra-cold atoms.


\section{II. System configuration and mathematical model}
 For our analysis, we will assume a quasi one-dimensional BEC, strongly trapped 
in the transverse directions ($x,y$) by a parabolic trap $V_\perp$  of characteristic frequency $\omega_\perp$ and weakly confined in the longitudinal 
dimension ($z$) by a shallow optical dipole trap with a Gaussian shape \cite{Stamper98,Martikainen99}. The choice of this geometry is justified 
by the need of outcoupling atoms along the $z$ axis, so the strength of the atomic interactions may be tuned to overcome the shallow potential.
Thus, the trap will have the following mathematical form:
\begin{equation}
V(\vec{r})=V_\perp+V_d=\frac{m\omega^2_\perp}{2} \left( x^2+y^2 \right)+V_0\left[1-\exp\left(-\frac{z^2}{L^2} \right)\right],
\label{fullpot}
\end{equation}
where $m$ is the mass of the atoms, $V_0$ is the depth of the shallow optical dipole potential and $L$ its characteristic width. The well-known 
mean field theory for a system of $N$ equal bosons of mass $m$, weakly interacting in a potential  yields a Gross-Pitaevskii \cite{GP} equation (GPE) of the form:
\begin{equation}
\label{GPE}
i \hbar \frac{\partial \Psi}{\partial t} = - \frac{\hbar^2}{2 m} \nabla^{2}\Psi 
+ V(\vec{r})\Psi + U(t,\vec{r})|\Psi|^2 \Psi,
\end{equation}
where $\Psi$ is the condensate wavefunction and $N = \int |\Psi|^2 \ d^3 \vec{r}$
the number of atoms. The 
coefficient $U(t,\vec{r}) = 4 \pi \hbar^2 a(t,z)/m$ characterizes the 2-body interaction which we 
consider a time and space-dependent function. In this situation, we can describe the dynamics of 
the condensate in the quasi-one dimensional limit as given by a factorized wavefunction 
of the form \cite{Perezgarcia98} $\Psi(t,\vec{r})=e^{-i\,\omega_\perp t}\Phi_0(x,y)\cdot\psi(t,z)$, 
where $\Phi_0(x,y)$ is a gaussian and $\psi(t,z)$ is determined by:
\begin{equation}
\label{NLSE}
i\frac{\partial \psi}{\partial \tau} = - \frac{1}{2}
\frac{\partial^2\psi}{\partial \eta^2} + f(\eta)\psi + g (\tau,\eta)|\psi|^2 \psi
\end{equation}
Our analysis is based on this equation.
We have introduced
 the dimension-less variables
 \begin{equation}
 \tau=\omega_\perp t\,,\qquad \eta=z/ r_\perp= z/\sqrt{\hbar/m\omega_\perp}
 \end{equation}
 which are the time measured in units of the inverse of the 
transverse trapping frequency and the length along the $z-$axis 
expressed in units of $r_\perp$ --- which determines the size of the cloud in the transverse $(x,y)$ plane. The 
functions 
\begin{equation}
f(\eta) = \frac{ V_d}{(\hbar\omega_\perp)}\,,\qquad g(\tau,\eta) =\frac{2 a(\tau,\eta)}{r_\perp}
\end{equation}
correspond, respectively,
to the trap and  to
the effective atomic interaction coefficient.
We must stress that $g$ depends on time and space and can be externally tuned;
 this is a key point of our analysis.
The new normalization is $\int |\psi|^2 d\eta=N$. 

The  simulations and figures that follow have been made with standard
 experimental parameters corresponding to $^7$Li atoms.
 In particular, we will take $\omega_\perp=1kHz$, 
 $r_\perp=3\mu$m. The choice of $^7$Li has been motivated 
for previous well-known results on soliton formation \cite{solitons1,solitons2},
however we must stress that our results can be straightforwardly extrapolated 
to different values of $\omega_\perp$ and
to other atomic species provided that external control of the scattering length
can be achieved. 
Since our goal is to study manipulations via Feshbach resonance tuning, we will
also fix the parameters determining the external potential:
 $L=15r_\perp$, $V_0=\hbar\,\omega_\perp/4$ --- in the following, we will also use the
  definitions: $\tilde L=L/r_\perp$, $\tilde V_0= V_0/(\hbar \,\omega_\perp)$.


\section{III. Soliton pair emission with a time-dependent scattering length}

\begin{figure}[htb]
{\centering \resizebox*{0.545\columnwidth}{0.5\columnwidth}{\includegraphics{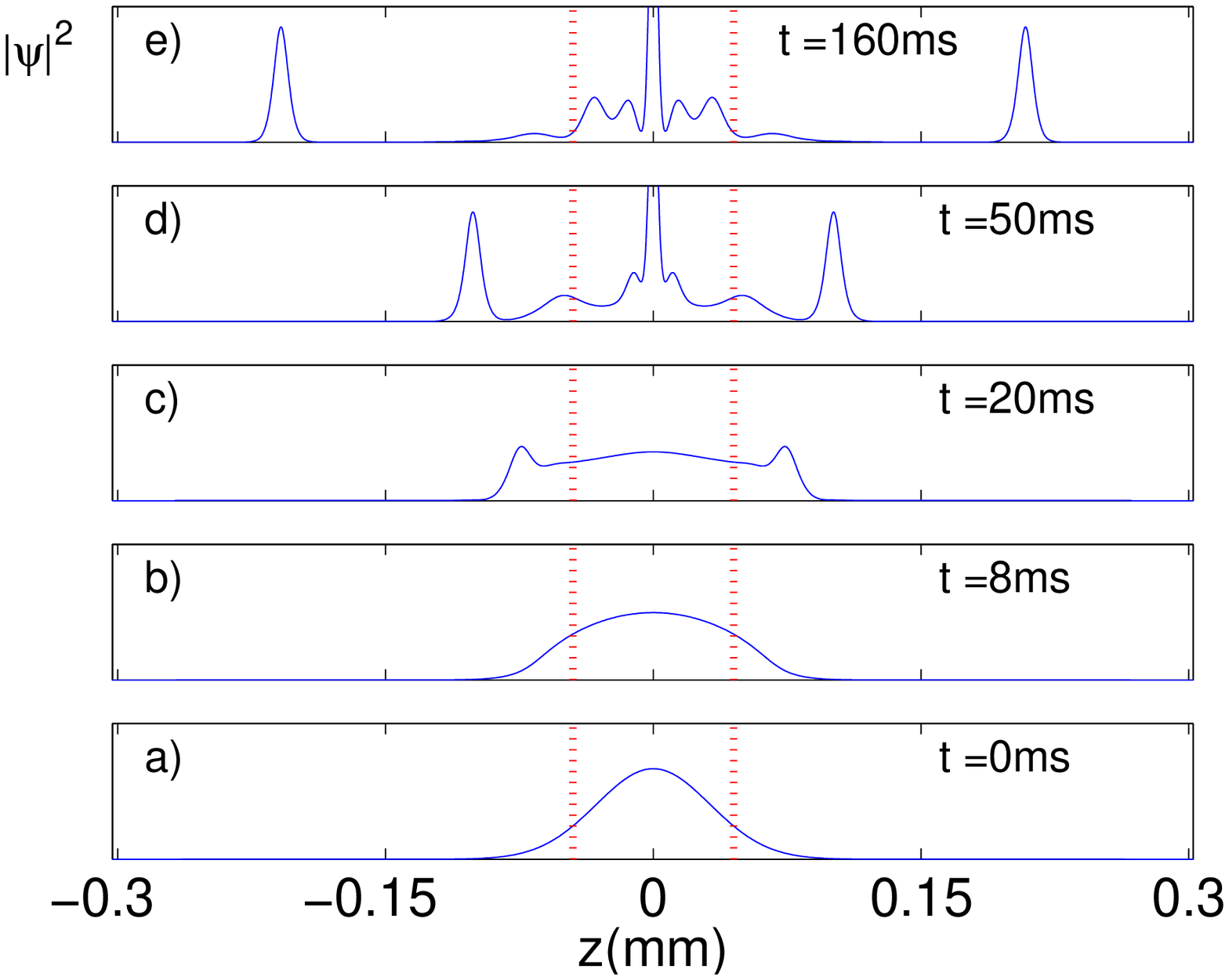}} 
\resizebox*{0.435\columnwidth}{0.5\columnwidth}{\includegraphics{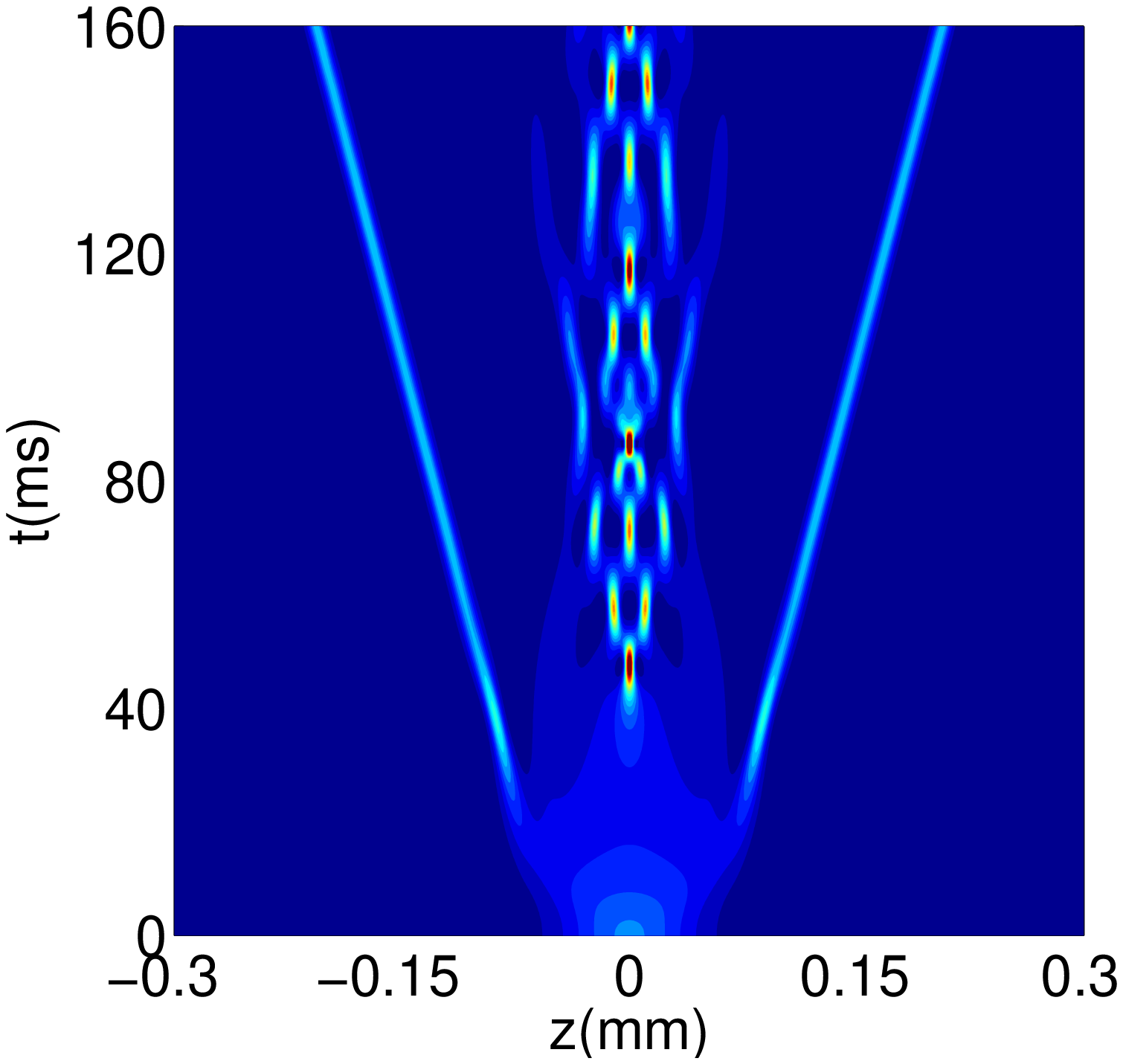}}\par}
\caption{(color online). 
Simulation showing the controlled emission
 by sharp tuning of the scattering length of a trapped BEC, 
yielding a pair of equal counter-propagating solitons. The experimental parameters correspond to $^7$Li
 atoms with  $N=5\times 10^4$, 
and $a_i=0.2$nm. 
At $t=0$, the scattering length is set to $a_1=1.5$nm and at
$t=t_s=8$ms, it is tuned to $a_2=-0.2$nm.
Left: evolution of the wavefunction 
$|\psi|^2(z)$ at different times indicated in the plots. 
The red dotted lines indicate the size of the trap.
Each of the two emitted solitons contains around 8500 atoms.
 Right: color contour map of the whole process. 
}
\label{fig1}
\end{figure}

In Fig. \ref{fig1} a numerical simulation of the effect of a sharp tuning with 
time of the scattering length $a$ is shown. For $\tau<0$, we have
$a=a_i= L\,\tilde V_0 \sqrt{\pi}/(2N)$, such that, in the Thomas-Fermi approximation,
there is a stationary solution with a Gaussian shape. This provides an initial condition
for the subsequent time evolution:
\begin{equation}
\psi(\tau=0,\eta)=\frac{\sqrt{N}}{\pi^\frac14 \sqrt{\tilde L}}
\exp\left(-\frac{\eta^2}{2\tilde L^2}\right)
\end{equation}
The value of $a$
 is assumed to change at $\tau=0$ to a larger one $a_1>a_i$. Therefore, the strength
of the repulsive atomic interactions will stretch the BEC wavefunction, as the shallow Gaussian potential is not strong enough to keep the cloud trapped. 
The broadening effect can be appreciated in the plots labelled a) and b) in the left row of Fig. \ref{fig1}. 
Once the cloud 
spreads out of the trap,
the interactions are instantaneously switched to a negative value $a_2$
at $t=t_s$. This can be done with standard 
magnetic \cite{FB1} or optical \cite{FB2} techniques which allow the values of $a$ to be tuned over a wide range. 
The effect on the cloud can be seen on the c) and d) plots in the left side of Fig. \ref{fig1}. The modulational instability effect \cite{Carr04} in the stretched 
condensate yields solitons that move away with a constant velocity as it can be appreciated in the contour plot from the right side of the figure.

\begin{figure}[htb]
{\centering \resizebox*{0.325\columnwidth}{0.45\columnwidth}{\includegraphics{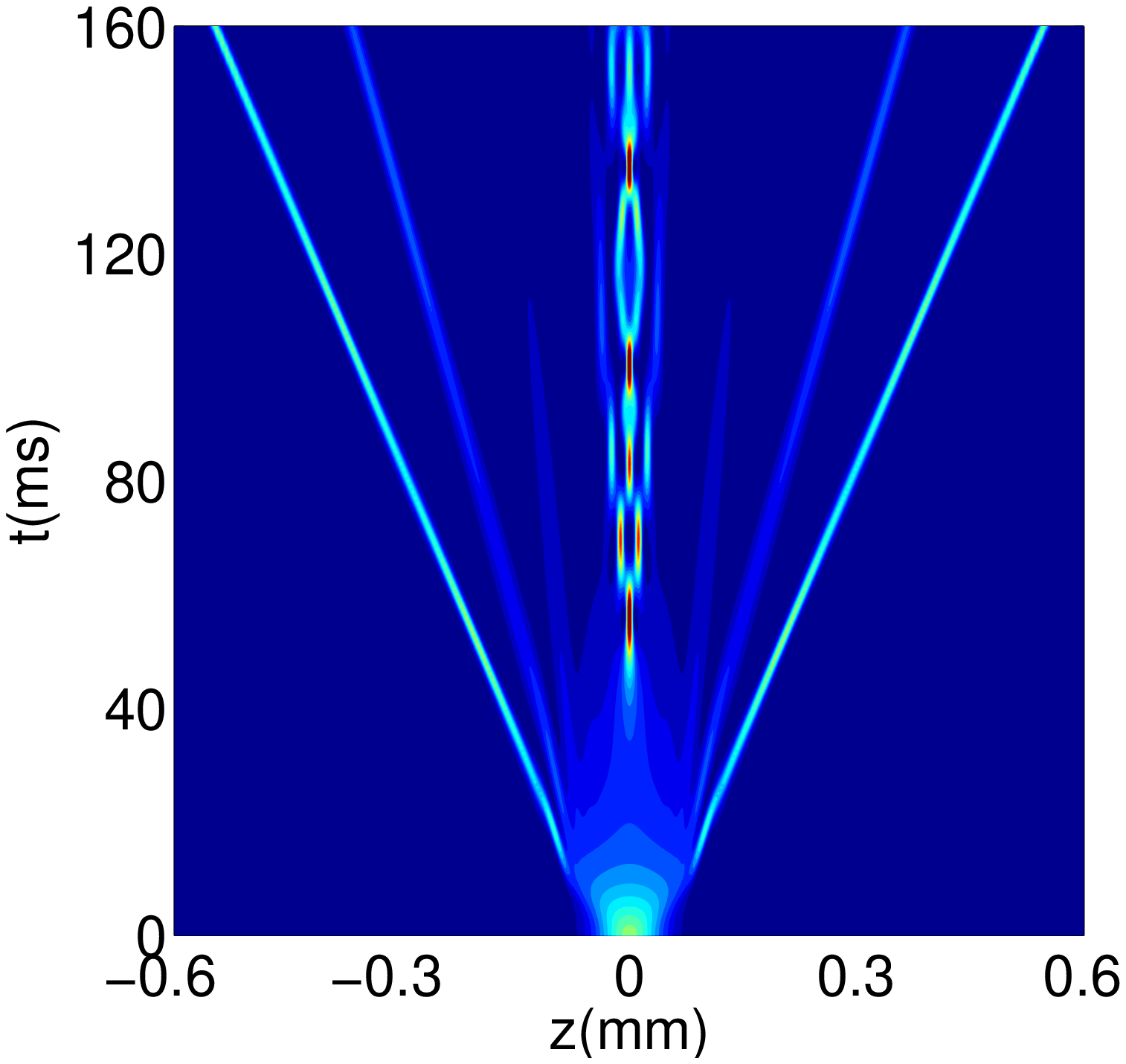}} 
\resizebox*{0.325\columnwidth}{0.45\columnwidth}{\includegraphics{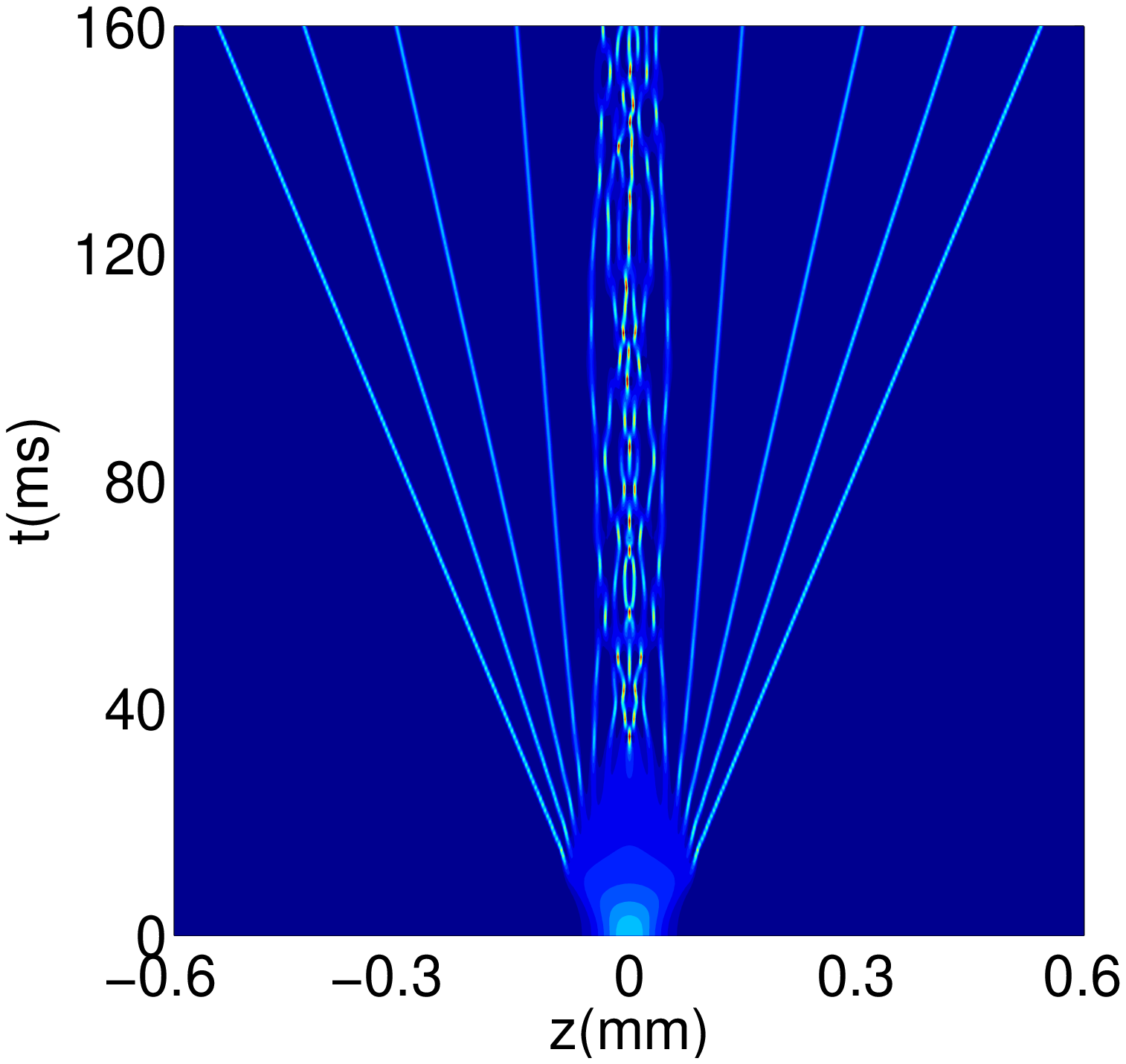}}
\resizebox*{0.325\columnwidth}{0.45\columnwidth}{\includegraphics{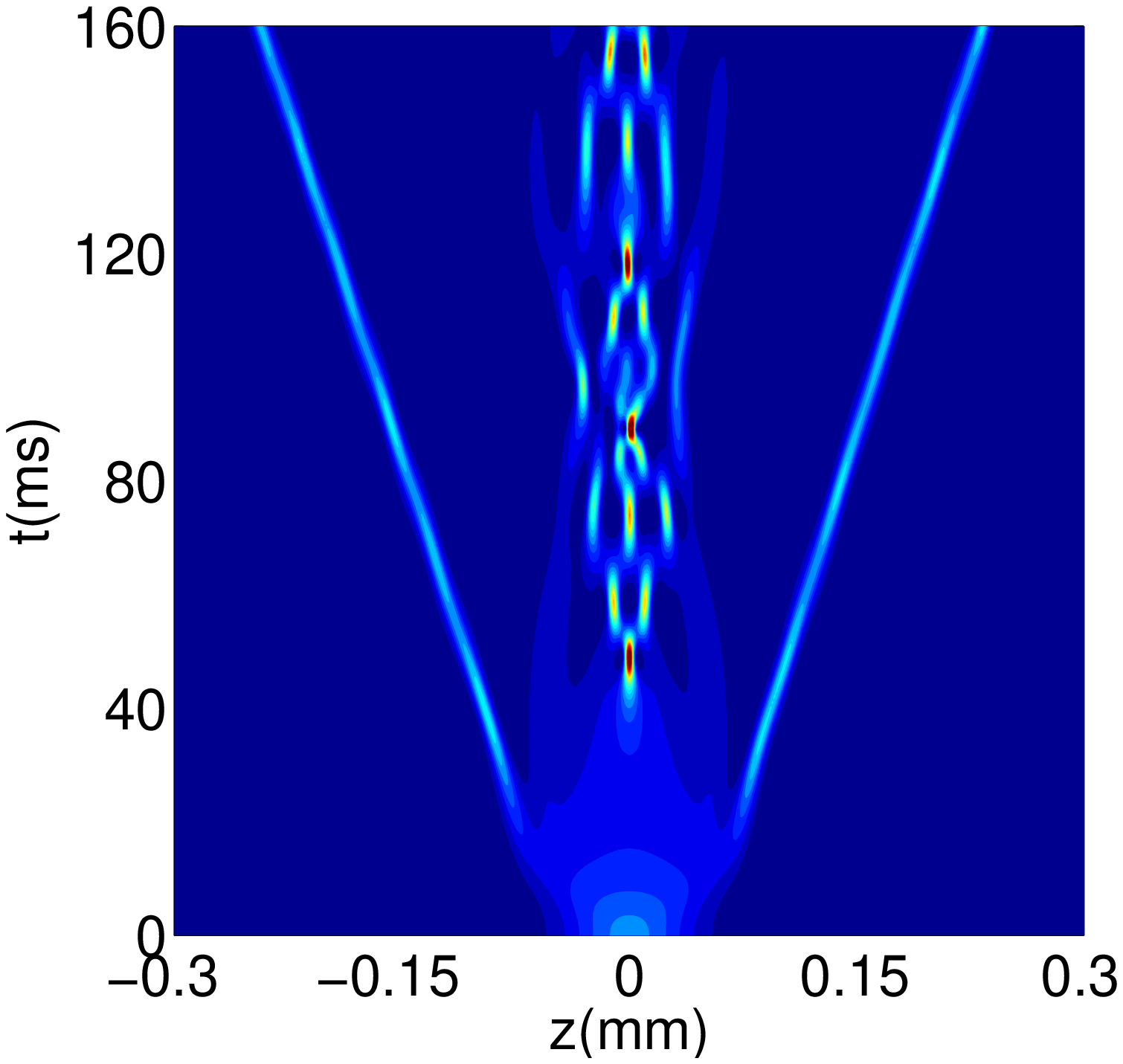}}\par}
\caption{(color online). In the first plot, we have taken $a_1=3$nm, $a_2=-0.2$nm.
In the second, $a_1=3$nm, $a_2=-1$nm. The total number of atoms
 $N=5\times 10^4$ is as in figure \ref{fig1}.
In the third plot, we repeat the simulation of Fig. \ref{fig1} but allowing 
noise in the
initial wave-function. We have discretized the initial function $\psi$, performed a
fast Fourier transform and multiplied its value at each point by a real pseudo-random
number obtained from a normal distribution with mean 1 and standard deviation 0.2.
After time evolution, the outcome does not change substantially.
}
\label{fig2}
\end{figure}

We must stress that not only a couple but a controllable number of soliton pairs can be produced by means of the technique proposed. This is
shown in Fig. \ref{fig2} where we illustrate the emission of an increasing number of pulses by an adequate control of the value of $a$ with time.
In the last plot of Fig. \ref{fig2}, we show that the process is not greatly modified in the 
presence of a limited amount of noise. We also stress that 
the production of
pairs can be achieved for a wide range of
 values of the scattering lengths. These observations accentuate the robustness of the process.

 
\begin{figure}[htb]
{\centering \resizebox*{0.325\columnwidth}{0.45\columnwidth}{\includegraphics{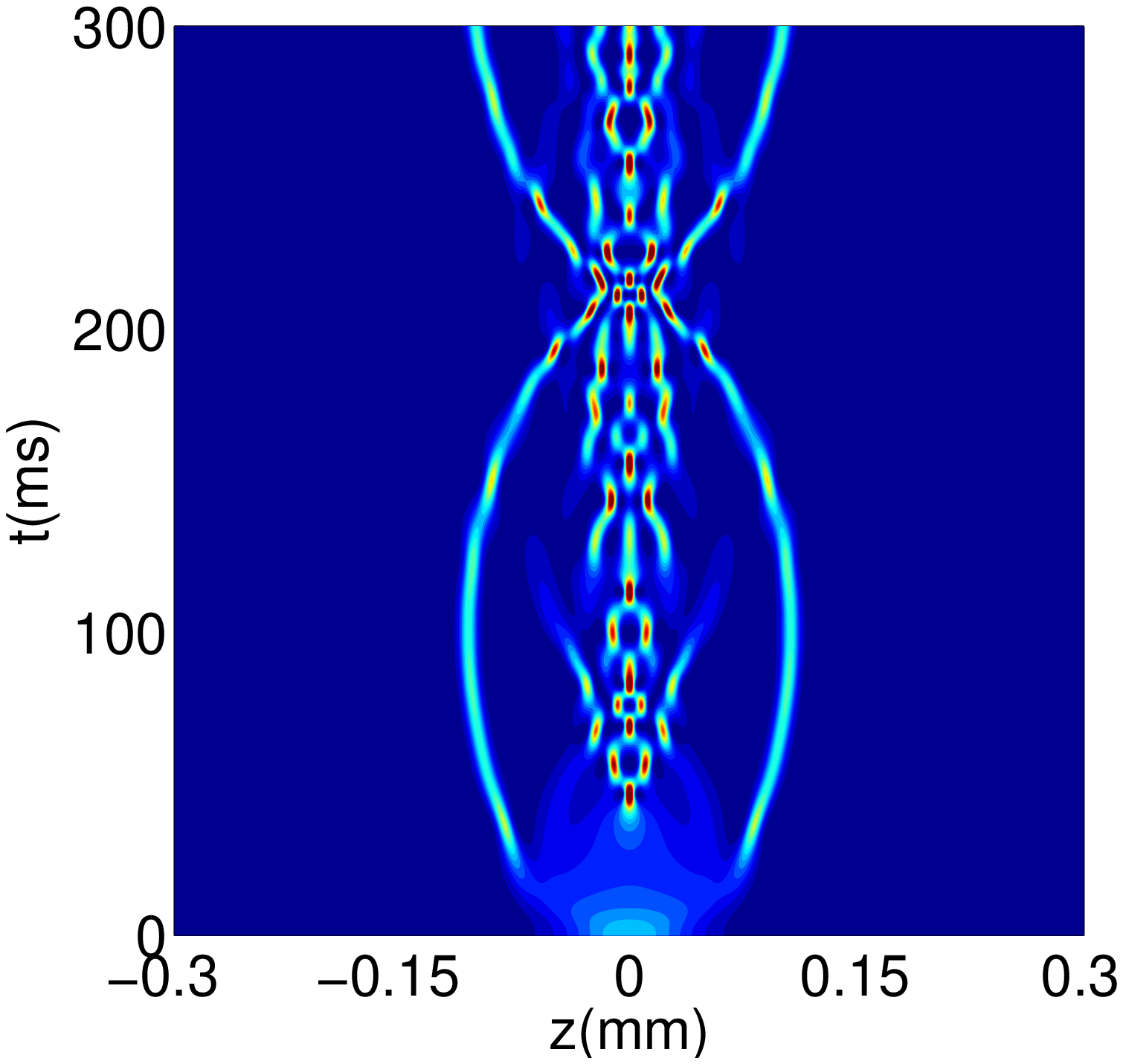}} 
 \resizebox*{0.325\columnwidth}{0.45\columnwidth}{\includegraphics{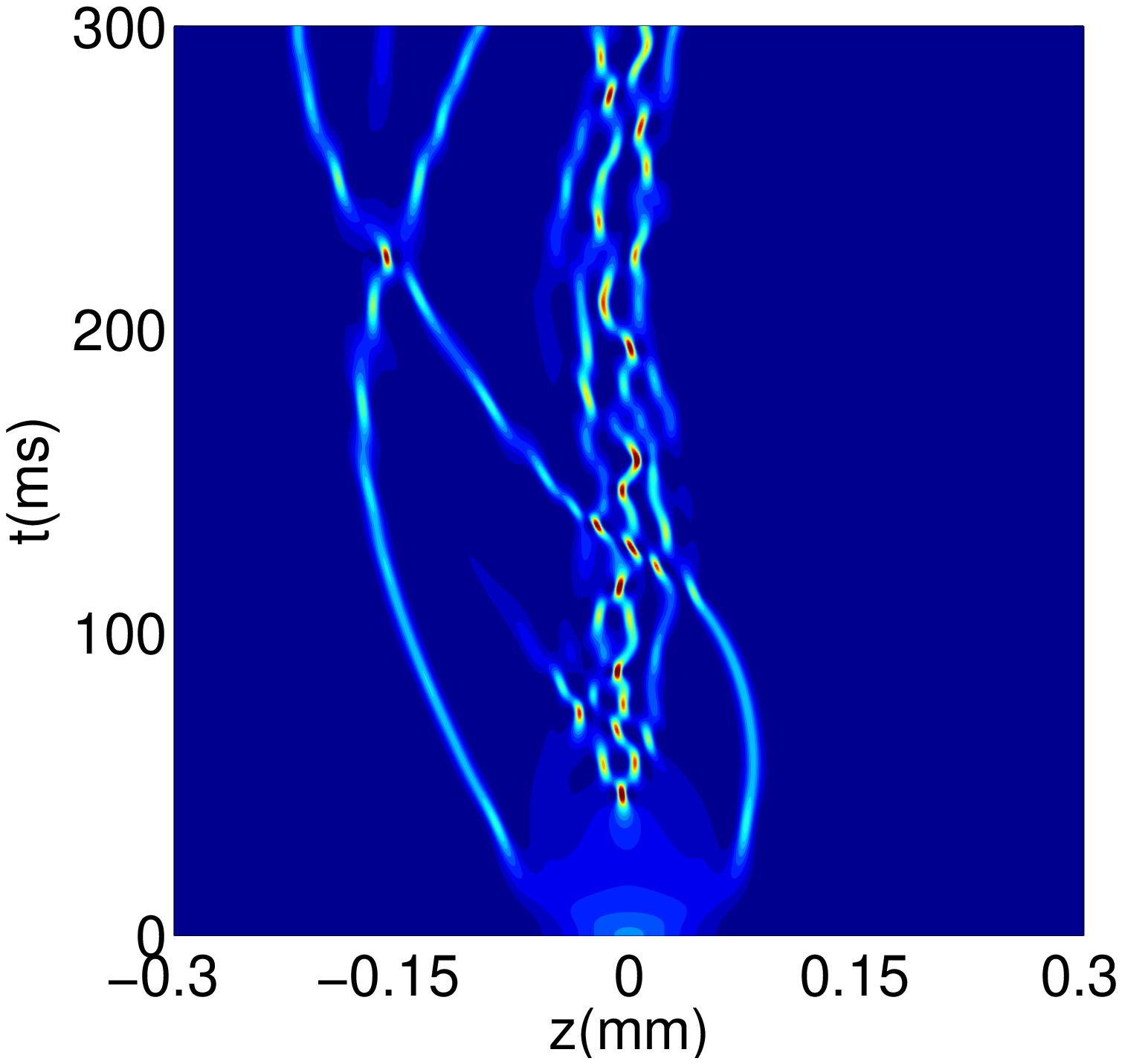}}
  \resizebox*{0.325\columnwidth}{0.45\columnwidth}{\includegraphics{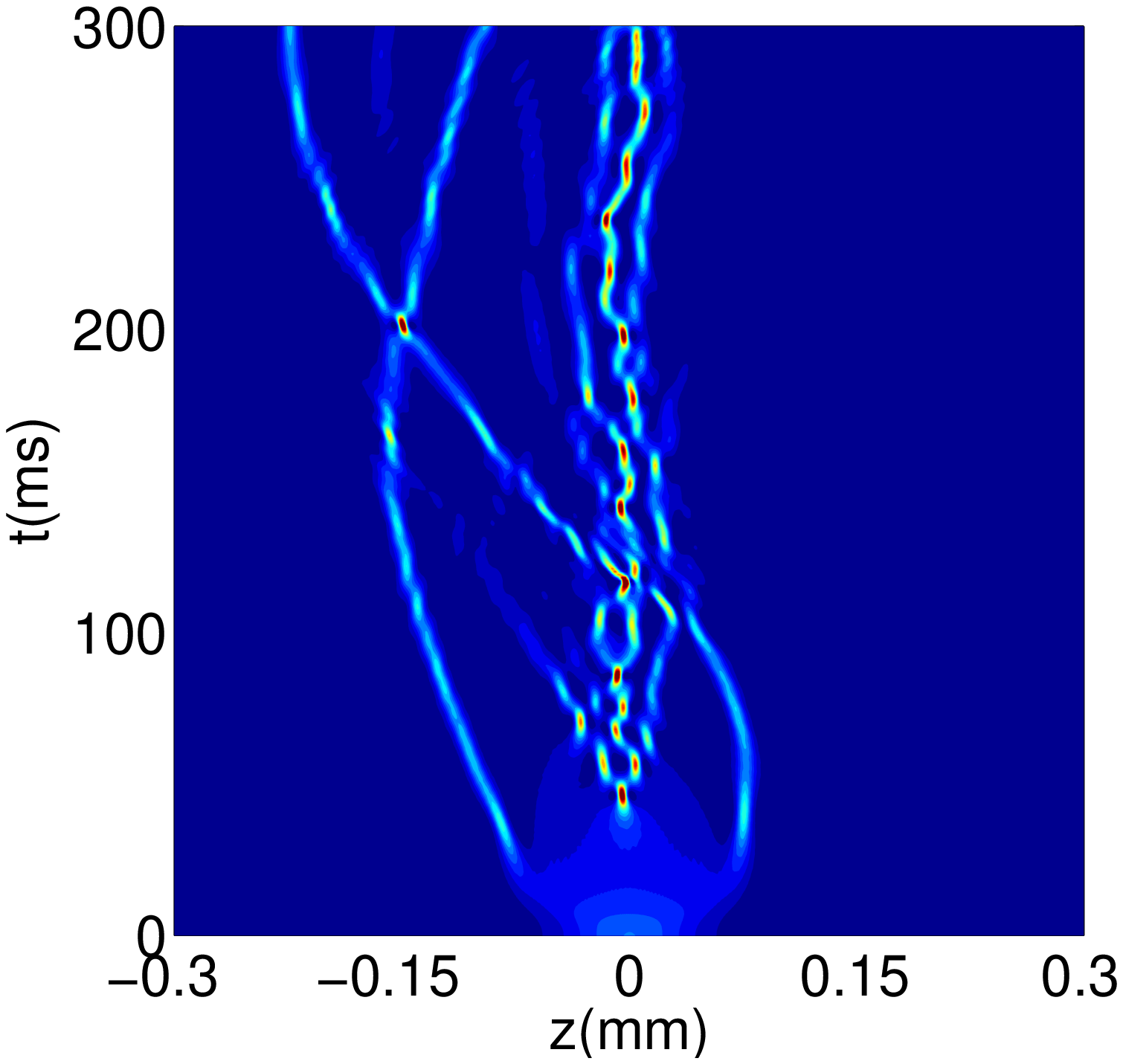}}
\par}
\caption{(color online). The first plot is a simulation with a harmonic potential along the
$z$-axis $\omega_z = 0.01 \omega_\perp$. 
The rest of
parameters are as in figure \ref{fig1}.
In the second plot we have added a small linear perturbation to
$V_d$, namely $0.0025 \hbar\,\omega_\perp\,z/r_\perp$.
The third plot is as the second one with noise 
included in the initial condition, as explained in the caption of Fig.
\ref{fig2}. }
\label{fig3}
\end{figure}

\begin{figure}[htb]
{\centering \resizebox*{0.545\columnwidth}{0.45\columnwidth}{\includegraphics{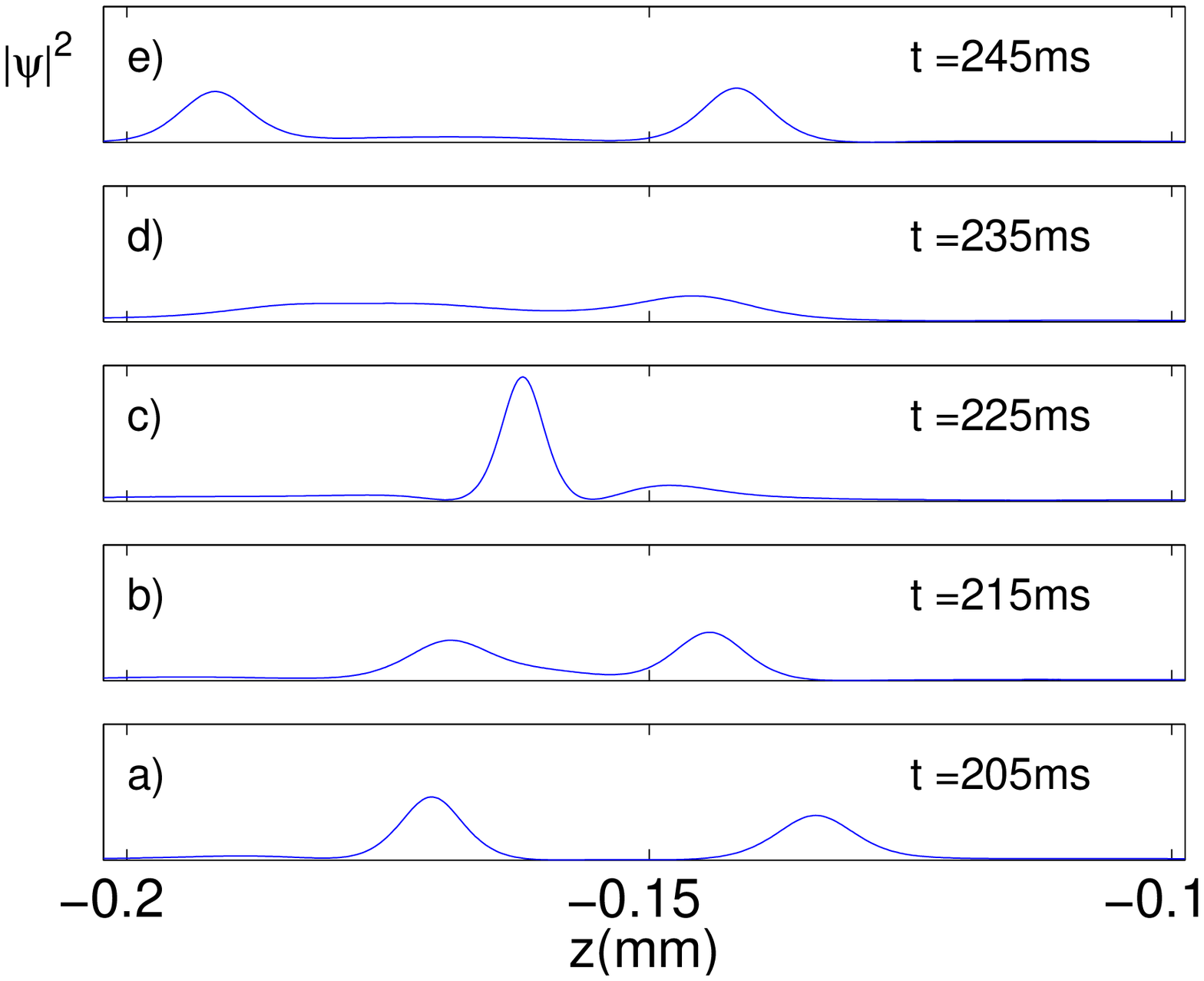}} 
\resizebox*{0.435\columnwidth}{0.45\columnwidth}{\includegraphics{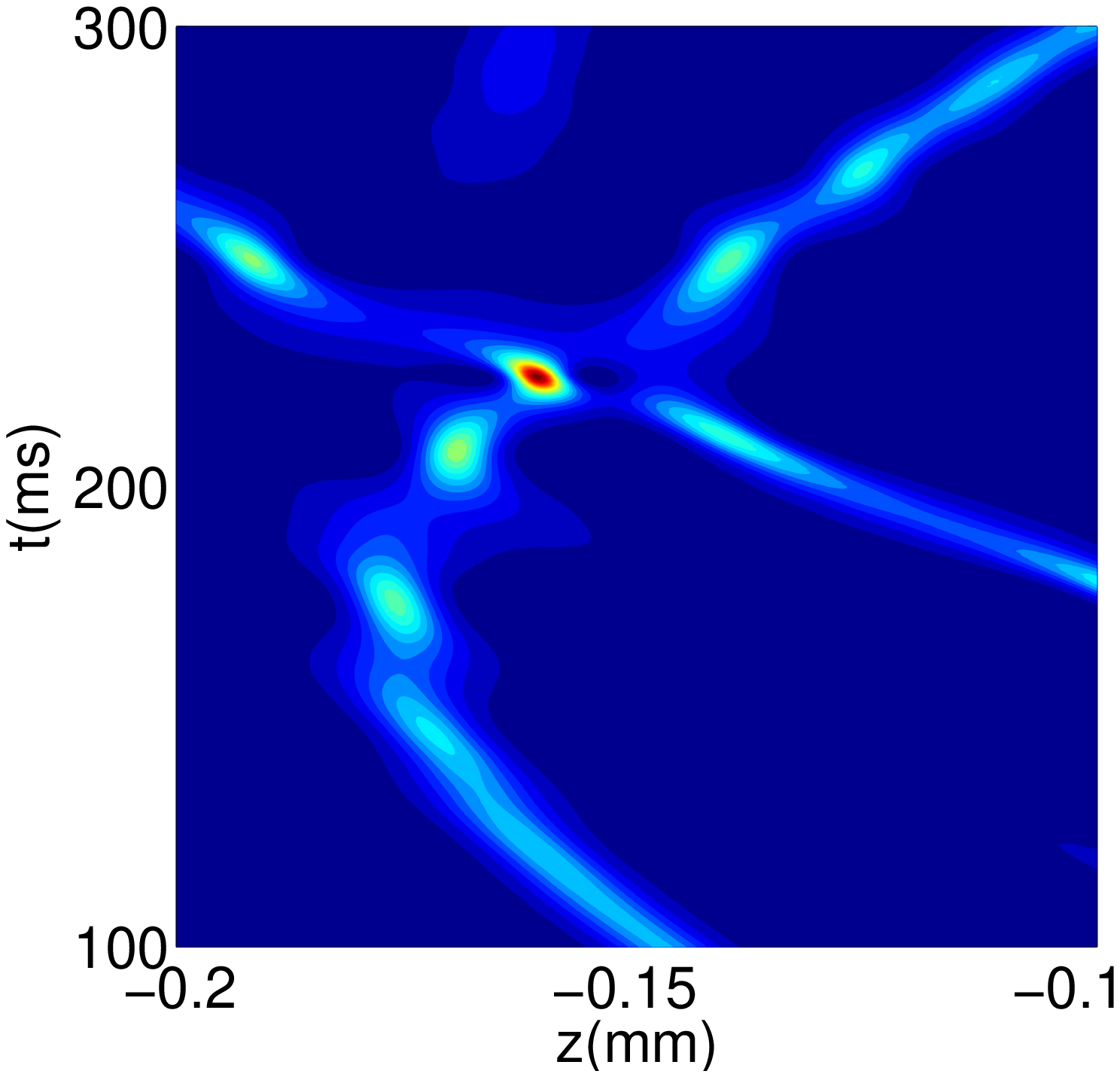}}\par}
\caption{(color online). A zoom of the second plot of figure \ref{fig3} around the region where 
the solitons cross.
Left: evolution of the wavefunction 
$|\psi|^2(z)$ at different times indicated in the plots. 
One can appreciate constructive-destructive interference when the solitons meet.
 Right: color contour map of the zoomed region.
}
\label{fig4}
\end{figure}

{\em Atomic Michelson interferometer.-} 
In Fig. \ref{fig3} we show a simulation of a simple Michelson interferometer that can be easily implemented
by simply adding a wide parabolic trap to 
the previous configuration, such that solitons eventually collide and yield an
interference pattern \cite{kumar}. In the absence of interaction of the soliton with the
rest of atoms the center of mass of the soliton would behave
 as a classical particle in the external potential due to an Ehrenfest theorem --- see for instance
\cite{moura}.
The time at which solitons meet and interfere would then be around $2\pi/\omega_z$.
In the figures one can appreciate that the interference happens earlier because of the attractive
interaction from the atoms in the trap.
We also show that if a linear perturbation is added to the potential, it is possible to have the solitons
interfering outside the Gaussian trap. In Fig. \ref{fig4}, we zoom in the interesting interference
region.


{\em Simple modelization.-}
In order to give the experimentalists a brief guide and get a qualitative
understanding of the physics, we can make a simple
 analysis of the process and give a rough approximation to
  the number of solitons produced
 as a function of the main physical parameters involved ($N$, $a_1$, $a_2$).

The first step of the process is the expansion due to large, positive $a_1$. We can find
an estimate of the expansion rate by making use of the 
variational method called averaged Lagrangian formalism
\cite{AL}. We use the following variational ansatz:
\begin{equation}
\psi = A(\tau) e^{-\frac{\eta^2}{2w(\tau)^2}}e^{i(\mu(\tau)+\eta\,\alpha(\tau)+\eta^2\,\beta(\tau))}
\end{equation}
where $A(\tau),w(\tau),\mu(\tau),\alpha(\tau),\beta(\tau)$ are real functions to be determined
by minimising the action from which GPE stems \cite{AL}. A straightforward analysis yields
$\alpha(\tau)=0$ as implied by the $\eta\to -\eta$ symmetry of the problem, 
$A(\tau) = \frac{\sqrt{N}}{\pi^\frac14 \sqrt{w(\tau)}}$ as required by normalization,
$\beta(\tau) = \frac{\dot w(\tau)}{2w(\tau)}$, a first order ODE for $\mu(\tau)$ which we
do not write and:
\begin{equation}
\ddot w(\tau) = - \frac{d\Pi}{dw(\tau)}
\label{eom1}
\end{equation}
with the pseudo-potential:
\begin{equation}
\Pi = \frac{1}{2w(\tau)^2}+\tilde L\,\tilde V_0 
\left(\frac{g}{\sqrt2 g_0 w(\tau)} - \frac{2}{\sqrt{\tilde L^2 + w(\tau)^2}} \right)
\label{pseudopot}
\end{equation}
We ought to solve (\ref{eom1}) with the initial conditions 
$w(0)= \tilde L\,,\ \dot w(0)=0$. Noticing that $\frac12 \dot w^2+\Pi$
is a conserved quantity, one can estimate the expansion velocity at the
time when the scattering length sign is swapped.
Considering that the second term in (\ref{pseudopot}) is the dominant one and
assuming that the change in $a$ is performed when $w \approx 2 \tilde L$, we find
$
\dot w|_{out} 
\approx (2/\pi)^{\frac14}\sqrt{a_1 N/L}
$.

\begin{figure}[htb]
{\centering  \resizebox*{1.05\columnwidth}{0.7\columnwidth}{\includegraphics{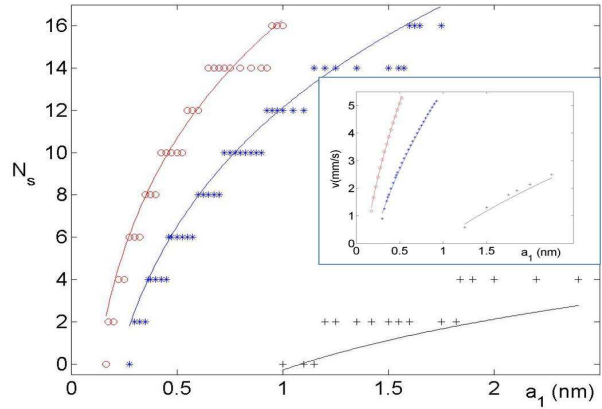}}}
\caption{Some examples of comparison of the number of emitted solitons (as computed from numerical
simulations) with the approximate expression (\ref{aproxim}) (solid lines). 
In each case we plot $N_s$ vs. $a_1$
for fixed $N$, $a_2$. The three curves, from top to bottom, correspond to $N=4.5\times 10^5$,
$a_2=-0.1$nm; $N=2.5\times 10^5$,
$a_2=-0.2$nm and
$N=0.5\times 10^5$,
$a_2=-0.2$nm. In the inset, we compare the speed of the fastest emitted solitons to fits
of the form $v\approx -b_1 + b_2 \sqrt{a_1 N}$ in the same three cases.
}
\label{figmodel}
\end{figure}

The second step of the process is soliton formation via modulational instability when
the scattering length becomes
$a_2 < 0$. For a flat initial wavefunction, there is a wavelength of the perturbation
 for which the instability is maximal \cite{modulational}, which, in terms of our
 dimension-less formalism reads
 $l = 2\pi \sqrt{ w/|g_2| N}$. The number of produced solitons is obtained by
 dividing the size of the wave-function by this length. We should just consider the atoms which
 are outside the trap so they can escape. Thus, the relevant condensate size for soliton formation
 can be approximated by $2 r_\perp( \dot w|_{out} \tau_{s} - \tilde L )  $.
 Putting everything together, we find the following expression for the functional
 dependence of the number of solitons
 on the adimensional parameters:
 \begin{equation}
N_s \approx 2\left[ \frac{ c_1 \sqrt{|g_2| N}(c_2\sqrt{g_1 N}
 - \tilde L)}{(g_1 N)^{1/4}}
\right]
\label{aproxim}
 \end{equation}
We have compared this estimate to the outcome of numerical simulations 
(keeping always $\tau_s=8$, $\tilde L=15$, $\tilde V_0=0.25$) and found that,
introducing the fitted coefficients $c_1=0.22$, $c_2=2.49$, 
Eq. (\ref{aproxim}) is a reasonably accurate approximation in a wide range of parameters, 
see Fig. \ref{figmodel}.
We have also confirmed by numerical simulations that the estimate for
$\dot w|_{out}$ is related to the velocity of the fastest outgoing soliton which
turns out to grow --- roughly --- as $\sqrt{a_1 N}$.

\section{IV. Soliton pair emission with a space-dependent scattering length}
\begin{figure}[htb]
{\centering \resizebox*{0.545\columnwidth}{0.45\columnwidth}{\includegraphics{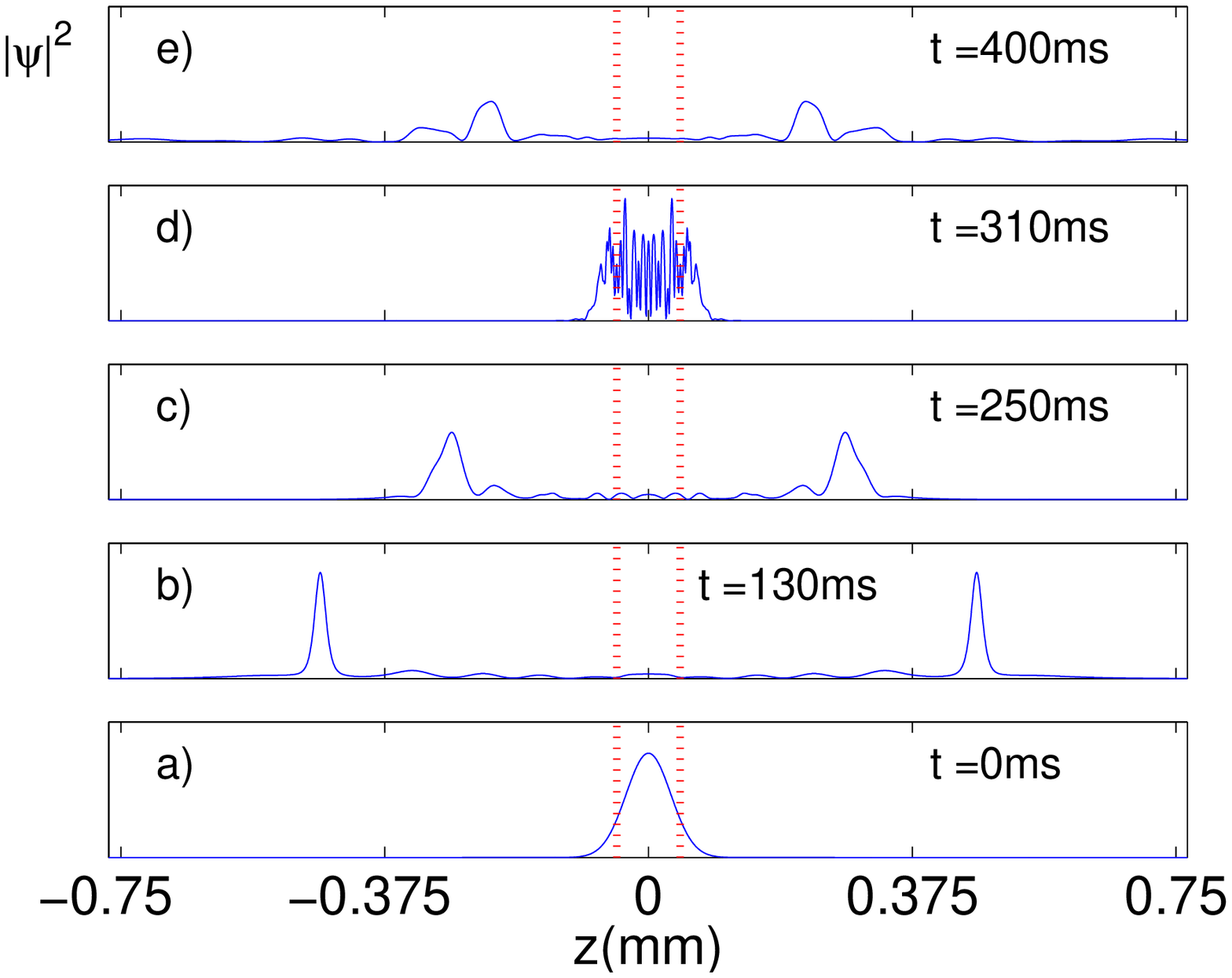}} 
\resizebox*{0.435\columnwidth}{0.45\columnwidth}{\includegraphics{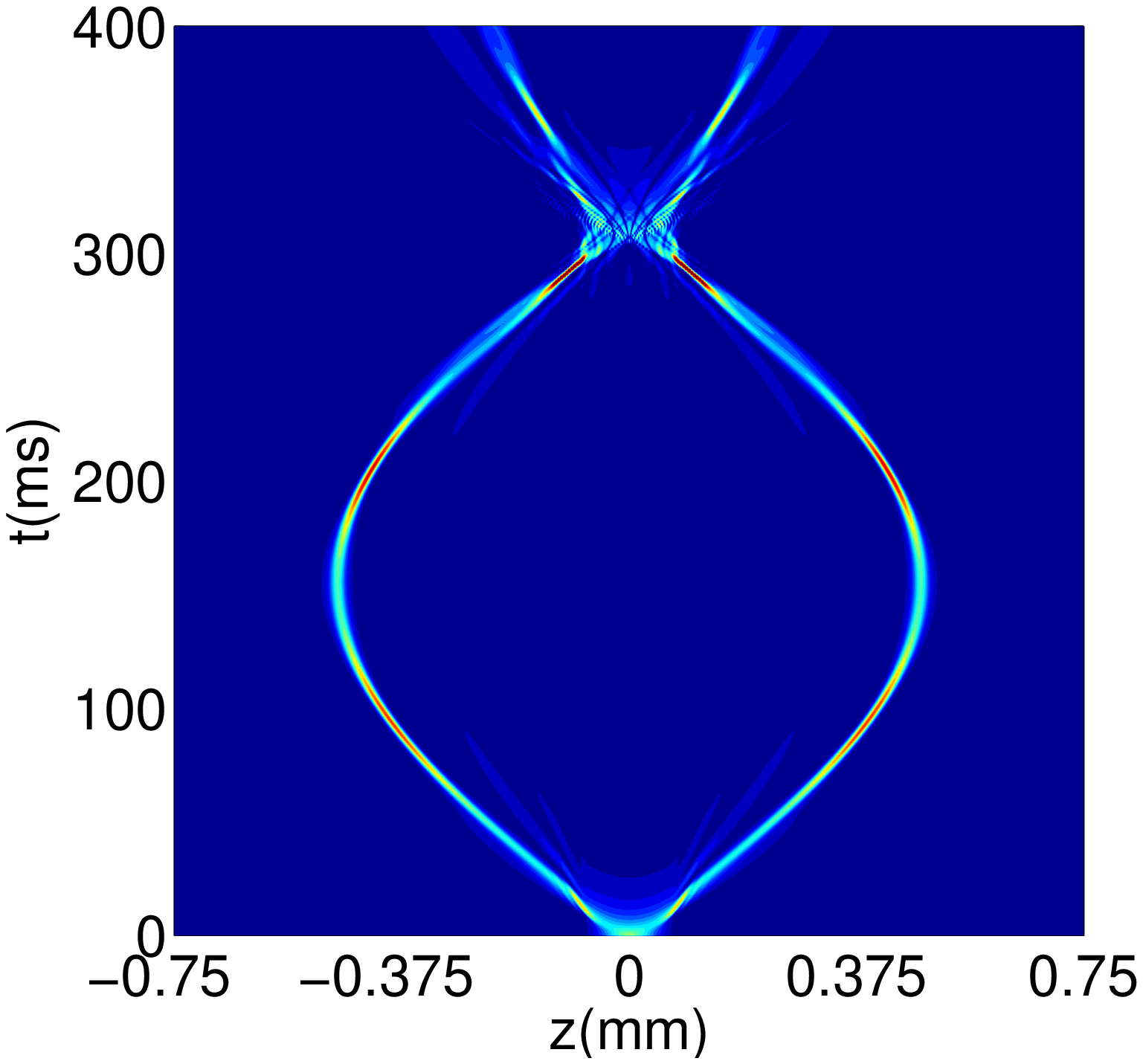}}\par
\hspace{1cm}
\hspace{1cm}
\resizebox*{0.8\columnwidth}{0.14\columnwidth}{\includegraphics{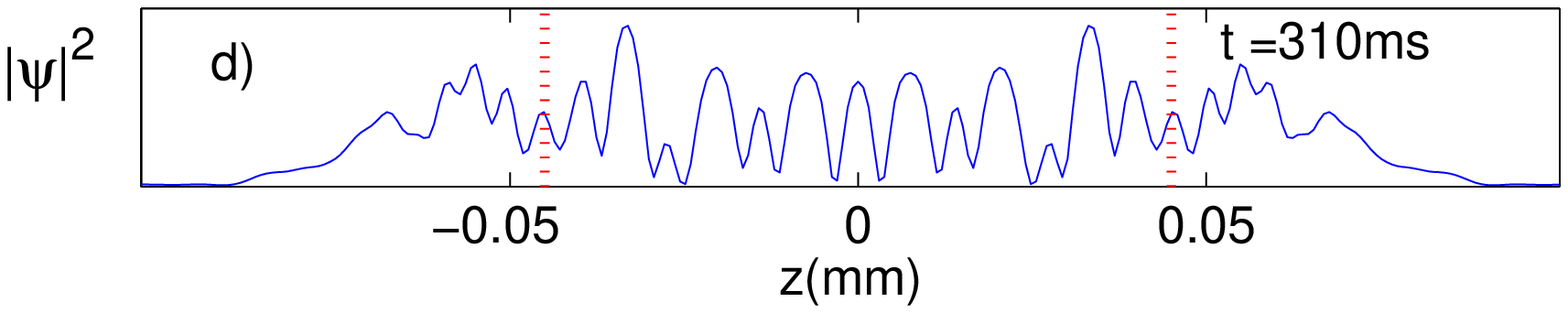}}\par
}
\caption{(color online). Emission and ulterior interference of a soliton pair
with a space-varying s-wave scattering length, Eq.
(\ref{alternative}). We have taken $N=3.5\times 10^4$, $a_1=5$nm,
$a_2=-0.1$nm. There is an
external harmonic potential $\omega_z=0.01 \omega_\perp$. In the graph below,
we show a zoom of plot d), exhibiting the interference pattern. }
\label{fig5}
\end{figure}

We explore now a second possibility and we consider an s-wave scattering length
which varies in space. We will utilize
\begin{equation}
a(z)=a_2 + (a_1-a_2)\exp(-z^2/L^2)
\label{alternative}
\end{equation}
for $\tau>0$.
As before $a_1 > a_i >0$, $a_2 < 0$. Since $a$ is positive for small
$z$, atoms are pushed out of the trap. The negative $a$ for larger $z$ contributes
to this stretching process and, additionally, it
 can re-pack the outgoing atom cloud into solitons \cite{rodas}.
We show in figure \ref{fig5} that, by appropriately tuning the physical parameters, 
it is possible to create a soliton pair while leaving the Gaussian trap almost empty.
This can be an advantage as compared to section III.
As in the previous case, these 
solitons can be made interfere by turning on a parabolic potential in the axial 
direction. It is worth noticing that, when the solitons re-enter the small $z$-region,
they disintegrate as they return to the region of positive $a$. The resulting atoms 
interfere yielding a typical pattern of fringes, see plot marked as d) in
figure \ref{fig5}.

\begin{figure}[htb]
{\centering \resizebox*{0.7\columnwidth}{0.45\columnwidth}{\includegraphics{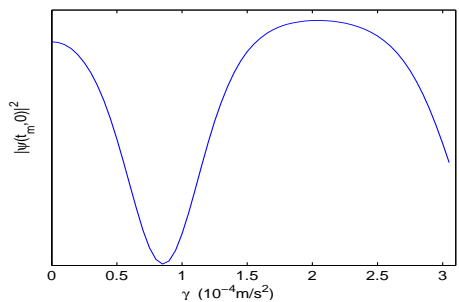}} 
}
\caption{The value of the atom density (arbitrary units) at the center of the trap at time
$t=310$ms, computed by numerical integration, as a function of the uniform acceleration
$\gamma$
caused by an external force.}
\label{fig7}
\end{figure}

As an example, let us discuss an extremely simple interferometric
Michelson-like
{\it gedanken}  experiment. We add a linear term to the external potential:
\begin{equation}
V_d = V_0 \left(1- \exp( - \frac{z^2}{ L^2} )\right)+
\frac12 m\,\omega_z^2 \,z^2 + m\,\gamma\,z
\end{equation}
We assume that $\gamma$ is tunable and that we wish to determine its value.
The linear potential affects the soliton trajectories and causes a 
displacement of the fringe pattern. As in a typical optical Michelson experiment,
the resolution of the interferometer should be related to the shift in $\gamma$ that
displaces one maximum of the interference pattern to the position of the next one.
A convenient way to estimate this resolution from numerical simulations is to
compute the atom density at the center of the trap 
when the interference
takes place, say at $t=t_m=310$ms and $z=0$, see Fig. \ref{fig7}. Since it is the whole
interference pattern being displaced, a similar plot would be found at any position
$z$ inside the trap.

The order of magnitude of the minimal acceleration that may be measured which such a device is related to the
spacing between maxima in the plot, namely $\gamma \approx 2 \times 10^{-4} $m/s$^2$.
We stress that this simple estimation is solely based on the computation of the displacement of the fringes.
When, eventually, a method is put forward to measure the fringe displacement in an actual experimental
set-up, there may well be experimental considerations
 that could influence the resolution.

Let us now provide an analytic estimation of the value of $\gamma$ found above.
The difference in the path of both parts of the beam when the interference pattern is
shifted by one maximum should be given by
$\frac{m\,v \Delta z }{\hbar}=\pi$ where $m\,v$ is the momentum when atoms interfere.
By considering the solitons as particles in a classical potential which depart initially from
$z=0$ with opposite velocities,
one can check that the difference in their paths before meeting again
is $\Delta z = 4 \gamma / \omega_z^2$.
In order to find a rough approximation to the momentum, we use the estimate of the previous section for the
velocity of the outgoing solitons, which in terms of dimensionful quantities reads
$v \approx \left(\frac{2}{\pi}\right)^\frac14 \sqrt{\frac{a_1 N}{L}}r_\perp \omega_\perp$.
Inserting these values of $\Delta z$ and $v$ in the condition above and inserting the value of the
parameters used in figures \ref{fig5}, \ref{fig7}, one finds 
$\gamma\approx \frac{\pi^{5/4}}{2^{9/4}}r_\perp \omega_z^2 \sqrt{
\frac{L}{a_1 N}}\approx 1.3\times 10^{-4}$m/s$^2$,
which captures the order of magnitude of the value obtained from the plot.


\section{V. Discussion}

We have shown that, by an appropriate tuning in time and/or space
of the s-wave scattering length related to the atom-atom interaction, it is 
possible to engineer a beam splitter for atomic interferometers. In particular,
we have analysed the possibility of generating a soliton pair.
We emphasize that this can be done in a controlled way and in a broad regime of
the physical parameters than can be typically achieved in BEC experiments.
We have shown the results for a particular set of dimensionful parameters, but
they can be easily generalised to different situations.
It is reasonable to expect that the robustness and lack of dispersion involved in 
soliton propagation might prove useful in ameliorating the precision obtained in
atom interferometers which deal with BEC but not with solitons. The next natural 
step is to design a concrete experiment to measure for instance the gravity acceleration
$g$ or to test gravity at short distances, but this lies beyond the scope of the present
work.

Our analysis of the dynamics has been performed using a reduction to one dimension of
the GPE, Eq. (\ref{NLSE}).
This 1-D approximation remains sensible as long as the atoms are not
energetic enough to get excited and probe the transverse directions.
We now discuss in more detail the validity and limitations of
Eq. (\ref{NLSE}).
The first relevant issue is how good the approximation leading to the reduction of
Eq. (\ref{GPE}) to its 1-D counterpart Eq. (\ref{NLSE}) is. For a recent analysis,
see \cite{billam2}. Of particular importance is the fact that, with attractive
interactions, collapses not describable with (\ref{NLSE}) may occur. 
Avoiding collapse in the transverse 2D space or collapse of a single soliton
require conditions \cite{Carr04} which in our notation read
$-g\,|\psi|^2 < 0.93$ and $-g\,N_{ais}< 1.254$, where $N_{ais}$ stands for the number
of atoms in a soliton. These limitations have, in part,
motivated our choice of physical parameters in the previous sections.

On the other hand, the GPE describes the evolution of a condensate
 at vanishing temperature. In any realistic
situation, there will be corrections controlled by the adimensional quantity $\frac{k_B T}{\hbar
\, \omega_\perp}$.
The thermal cloud can produce friction for the motion of the soliton \cite{sinha} or deplete the
condensation and even cause
incoherent splitting of a soliton after some time \cite{buljan}.
These considerations affect the coherent evolution of the system 
and therefore the 
sensibility of an eventual soliton interferometer will depend on the temperature
of the atom cloud. It would be of great interest to explore this point.

Moreover, when one deals with elongated condensates, as it is the case of this paper, 
one has to take into account that there may be non-negligible fluctuations of the phase along
the axial direction, 
even well below the BEC transition temperature $T_c$ \cite{petrov1}.
The most important contributions come from fluctuations of wavelength larger than $r_\perp$,
due to the similarity of the system to 1D trapped gases \cite{petrov2}.
The equilibrium state is a {\it quasi-condensate}. 
If these fluctuations were large, they would hinder any interferometric measurement so
it is essential to estimate their amplitude. Using the Thomas-Fermi approximation for
an elongated trap with repulsive interactions, in \cite{petrov1} it was shown that phase
fluctuations are controlled by the parameter:
\begin{equation}
\delta_L^2(T) = \frac{T}{T_c}\,\left(\frac{N}{N_0}\right)^{3/5}\delta_c^2\,
\end{equation}
where $N$ is the total number of atoms in the sample, $N_0$ the number of those
which are condensed and:
\begin{equation}
\delta_c^2 = \frac{16\, a^{2/5}\, m^{1/5}\, \omega_\perp^{22/15}}{15^{3/5}\, N^{4/5}\,\hbar^{1/5}
\,\omega_{\parallel}^{19/15}}
\label{deltac}
\end{equation}
where $\omega_{\parallel}$ corresponds to the trapping potential in the longitudinal axis.
In our set-up, the potential along $z$ given in Eq. (\ref{fullpot}) is only
parabolic in the vicinity of $z=0$, but we can write
$\omega_{\parallel}^2 \approx \frac{1}{m} \left[
\partial_z^2 V(\vec r)\right]_{x=y=z=0}= 2 \tilde V_0 \omega_\perp^2 / \tilde L^2$.
By inserting the numerical values describing the initial conditions 
of the simulations of figures \ref{fig1} and
\ref{fig5}, we find $\delta_c^2 = 0.18$ and  $\delta_c^2 = 0.23$, respectively.
Assuming $N\approx N_0$ and, obviously $T<T_c$, we see that fluctuations are not large
on any distance scale since $\delta_c^2 < 1$. 
This shows that for the situations we have analysed this effect would
not spoil an eventual interferometric measurement, even if it may reduce the contrast of the
observable signatures. If in some other case one had $N\approx N_0$ but $\delta_c^2 \gg 1$,
it would be necessary to cool down the condensate below $T_\phi = T_c /\delta_c^2$ 
\cite{petrov1} in order
to avoid problems with this kind of phase fluctuations.

Finally, we would like to mention that soliton splitting in different
situations has been investigated in
a quantum framework beyond mean-field GPE in \cite{castin,streltsov,martin}.

\

\centerline{{\bf Acknowledgements}}

This work was supported by Xunta de Galicia (project 10PXIB383191PR) and by the
Univ. Vigo research programme.
The work of A. Paredes is supported by the Ram\'on y Cajal programme. 
M.M. Valado gratefully acknowledges funding by COHERENCE Network and INO-CNR. 
The work of D. Feijoo is supported by a research grant of Univ. Vigo.



\begin{thebibliography}{99}

\bibitem{Anderson95}{M. H. Anderson, {\em et al.},
Science {\bf 269}, 198 (1995);
K. B. Davis, {\em et al.}, Phys. Rev. Lett. {\bf 75}, 3969 (1995).}
                                                                                                                   
\bibitem{Mewes97}{ M. -O. Mewes, {\em et al.},
Phys. Rev. Lett. {\bf 78}, 582 (1997).}

\bibitem{RMP}{ A. D. Cronin, J. Schmiedmayer, D. E. Pritchard,
Rev. Mod. Phys. {\bf 81}, 1051-1129 (2009) .}

\bibitem{sensors}{B. Lucke, {et al.}, Science {\bf 334}, 773 (2011).}

\bibitem{precision}{A. Widera, {et al.}, Phys. Rev. Lett. {\bf 92}, 160406 (2004). }

\bibitem{gravitation}{S. Dimopoulos, P.W. Graham, J.M. Hogan,
M.A. Kasevich, S. Rajendran, Phys. Rev. D {\bf 78}, 122002 (2008). }

\bibitem{interferometer}{ Y. Shin, {et al.}, Phys. Rev. Lett. {\bf 92}, 050405 (2004); T. Schumm, {\em et al.}, Nature Phys. {\bf 1}, 57 (2005). }

\bibitem{hyllus} J. Chwede\'nczuk, P. Hyllus, F. Piazza, A. Smerzi,
arxiv:1108.2785.



\bibitem{gross}{C. Gross, {et al.}, Nature {\bf 464}, 1165 (2010). }

\bibitem{solitons1} {K. E. Strecker, G.B. Partridge, A.G. Truscott R.G. Hulet,
 Nature {\bf 417}, 150 (2002).
 
 \bibitem{solitons2} L. Khaykovich, {\em et al.}, Science {\bf 296}, 1290 (2002).}


\bibitem{parker} N.G. Parker, A.M. Martin, S.L. Cornish, C.S. Adams,
J. Phys. B: At. Mol. Opt. Phys. {\bf 41} (2008) 045303.



\bibitem{castin} C. Weiss, Y. Castin, Phys. Rev. Lett. {\bf 102},
010403 (2009).

\bibitem{streltsov} A.I. Streltsov, O.E. Alon, L.S. Cederbaum,
Phys. Rev. A {\bf 80}, 043616 (2009).

\bibitem{billam} T.P. Billam, S.L. Cornish, S.A. Gardiner,
Phys. Rev. A {\bf 83}, 041602 (2011).

\bibitem{stoof} U. Al Khawaja, H.T.C. Stoof, 
New J. Phys. {\bf 13}, 085003 (2011).

\bibitem{martin} A.D. Martin, J. Ruostekoski,
New J. Phys. {\bf 14}, 043040 (2012).

\bibitem{helm} J.L. Helm, T.P. Billam, S.A. Gardiner, 
arxiv:1203.3080.



\bibitem{Carr04}{L. D. Carr and J. Brand,
Phys. Rev. A {\bf 70}, 033607 (2004)}.

\bibitem{chip}{ H. Ott, J. Fortagh, G. Schlotterbeck, A. Grossmann, C. Zimmermann,
Phys. Rev. Lett. {\bf 87}, 230401 (2001);
W. H\"ansel, P. Hommelhoff, T. W. H\"ansch, J. Reichel , Nature {\bf 413}, 498 (2001).
}

\bibitem{FB1}{S. Inouye, {\em et al.},
Nature {\bf 392}, 151 (1998).}

\bibitem{FB2}{M. Theis,  \emph{et al.},Phys. Rev. Lett. 93, 123001 (2004).}

\bibitem{carrbrand2} {L. D. Carr and J. Brand,
Phys. Rev. Lett. 92, 040401 (2004).}

\bibitem{Stamper98}{ D. M. Stamper-Kurn, {\em et al.},
Phys. Rev. Lett. {\bf 80}, 2027 (1998).}

\bibitem{Martikainen99}{J. P. Martikainen, Phys. Rev. A {\bf 63}, 043602 (2001).}

\bibitem{GP}{E. P. Gross, Il Nuovo Cimento {\bf 20}, 454-457 (1961); L.P. Pitaevskii, Soviet Physics JETP {\bf 13} 451-454 (1961);
F. Dalfovo, S. Giorgini, L.P. Pitaevskii, S. Stringari, Rev. Mod. Phys. {\bf 71}, 463 (1999).}



\bibitem{Perezgarcia98}{V. M. P\'erez-Garcia, H. Michinel, and H. Herrero,
Phys. Rev. A {\bf 57}, 3837 (1998).}


\bibitem{kumar} {V.Ramesh Kumar, R. Radha, P.K. Panigrahi,
Phys. Lett. A 373 (2009), 4381-4385. 
}


\bibitem{moura} M.A. de Moura, J. Phys. A: Math. Gen. 27 (1994), 7157.


\bibitem{AL}{V. M. P\'erez-Garc\'{\i}a, H. Michinel, J.I. Cirac,
M. Lewenstein, P. Zoller, Phys Rev Lett. \textbf{77} 5320 (1996).}



\bibitem{modulational} {V.I. Bespalov and V.I. Talanov},
JETP Letters-USSR  3-12 (1966), 307.

\bibitem{rodas} M.I. Rodas-Verde, H. Michinel, V.M. P\'erez-Garc\'\i a,
 Phys. Rev. Lett. \textbf{95}, 153903 (2005).





\bibitem{billam2} T.P. Billam, S.A. Wrathmall, S.A. Gardiner,
Phys. Rev. A {\bf 85}, 013627 (2012).



\bibitem{sinha} S. Sinha, A.Y. Cherny, D. Kovrizhin, J. Brand, Phys. Rev. Lett.
{\bf 96}, 030406 (2006).

\bibitem{buljan} H. Buljan, M. Segev, A. Vardi, Phys. Rev. Lett, {\bf 95},
180401 (2005).

\bibitem{petrov1}  D.S. Petrov, G.V. Shlyapnikov, J.T.M. Walraven, Phys. Rev. Lett, {\bf 87},
050404 (2001).

\bibitem{petrov2}  D.S. Petrov, G.V. Shlyapnikov, J.T.M. Walraven, Phys. Rev. Lett, {\bf 85},
3745 (2000).






























\end{thebibliography}
\end{document}